\documentclass{article}
\def\sbd{\vspace{8pt}\noindent\bf}

\begin{document}
\begin{center}
{\Large{\bf A Unification Algorithm for Second-Order Linear Terms}}
\\[10pt]
{\bf Gilles Dowek}\\[10pt]
{\bf INRIA}\def\thefootnote{\fnsymbol{footnote}}\footnote[1]{B.P. 105, 78153 Le Chesnay CEDEX, France. Gilles.Dowek@inria.fr}
\def\thefootnote{\arabic{footnote}}
\\[15pt]
\end{center}
\thispagestyle{empty}

We give an algorithm for the class of second order unification
problems in which second order variables have at most one occurrence.
As an application we give another algorithm that generalizes both first order
unification and second order matching.

\section{An Algorithm for Second Order Linear Unification Problems}

We consider simply typed lambda-calculus with $\alpha$, $\beta$ and
$\eta$ rule. Variables and constants are restricted to be at most
second order. A {\it unification problem $N$} is a set
of equations $a = b$ such that $a$ and $b$ have the same type $T$ and
this type is first order.
A {\it solution} of a unification problem $N$ is a substitution $\sigma$ such
that for each equation $a = b$ of $N$, 
$\sigma a =_{\alpha \beta \eta} \sigma b$.

{\sbd Remark:} Normal terms of a first order type are atomic, i.e. they
have the form $(f~a_{1}~...~a_{n})$, where $f$ is a variable or a
constant. Such a term is said to be {\it flexible}
when $f$ is a variable, it is said to be {\it rigid} when $f$ is a
constant.

A second order unification problem is said to be {\it linear} when
each second order variable has at most one occurrence in the problem
(without any restriction on the number of occurrences of constants and
first order variables). 

A second order unification problem is said to be {\it
superficial} if all the equations of this problem have the form
$(F~a_{1}~...~a_{n}) = (G~b_{1}~...~b_{p})$ where all the variables
that occur in the terms $a_{1}, ..., a_{n}, b_{1}, ..., b_{p}$ are
first order, i.e. second order variables can only occur as head
symbols.

{\sbd Proposition:} There exists an algorithm that transforms each
second order unification problem $N$ into a second order superficial
unification problem which is equivalent to $N$. 

While there is an equation of the form
$$(F~...~a~...) = t$$
where $F$ is a second order variable or constant, $t$ a term, 
and $a$ a term which is not a variable, replace this equation by
$$(F~...~x~...) = t$$
$$x = a$$
where $x$ is a fresh first order variable. 
This process obviously terminates.

{\sbd Corollary:} (Amiot \cite{Amiot}) There is no decision algorithm 
for second order superficial problems.

{\sbd Remark:} If the problem we start with is linear then so is the 
superficial problem built from it.

{\sbd Proposition:}
There exists an algorithm that solves second order superficial linear 
unification problems.

Let us consider equations between a flexible term and a rigid term of
a second order superficial problem. If the head variable of the
flexible term is first order then the equation has the form 
$$x = b$$
where $x$ is a first order variable and all the variables occurring in
the term $b$ are first order.
If the head variable of the flexible term is second order 
then the equation has the form
$$(f~a_{1}~...~a_{n}) = b$$
where $f$ is a second order variable and all the variables occurring
in the terms $a_{i}$ and $b$ are first order.

We consider a variant of the unification algorithm of \cite{Huet75}
\cite{Huet76} in which when we consider an equation on the form 
$x = b$ such that all the variables occurring in the term $b$ are first order,
if the variable $x$ does not occurs in $b$ then one substitutes $b$
for $x$ and otherwise one fails. When restricted to second order superficial 
unification problems, the projection and imitation substitutions are 
considered in this algorithm only when the head variable of the
flexible term is second order.

{\sbd Lemma:} The unification tree of a second order superficial
linear unification problem is finite.

Le $N$ be a problem. Let $v(N)$ be the number of distinct first order variables
occurring in $N$.
Let $W(N) = \{t_{1}, ..., t_{n}\}$ be the multiset of rigid terms
of equations between a flexible and a rigid term such that the
head variable of the flexible term is second order.
Let $w(N) = |t_{1}| + ... + |t_{n}|$, where
$|t|$ is the number of variable occurrences in the term $t$.

Let $N$ be a second order superficial linear unification problem
and $N'$ a son of this problem in the unification tree. Then
$(v(N'),w(N')) < (v(N),w(N))$ 
for the lexicographic order on ${\it \omega}^{2}$.
\begin{itemize}
\item
When we instantiate a first order variable $v(N') < v(N)$.
\item
When we perform a {\it projection} substitution, a term gets out from
$W(N)$, $w(N') < w(N)$ and\\ $v(N') = v(N)$.
\item 
When we perform an {\it imitation} substitution, we replace the equation
$$(f~a_{1}~...~a_{n}) = (\Phi~b_{1}~...~b_{p})$$
by 
$$(h_{1}~a_{1}~...~a_{n}) = b_{1}$$
$$...$$
$$(h_{p}~a_{1}~...~a_{n}) = b_{p}$$
The term $(\Phi~b_{1}~...~b_{p})$ gets out form $W(N)$ 
and the terms $b_{1}, ..., b_{p}$ get in,
$w(N') < w(N)$ and\\ $v(N') = v(N)$.
\end{itemize}

{\sbd Proposition:} There exists an algorithm that solves second order linear
unification problems.

{\sbd Remark:}
In the algorithm above, both variables and constants are restricted to
be at most second order and we consider only equations between terms
of a first order type. A similar algorithm can be designed for the
case in which constants and bound variables are at most third order
and equations relate terms of an arbitrary type.

The main problem is one cannot transform the problem 
$[x_{1}:T_{1}] ... [x_{n}:T_{n}](F~...~a~...) = t$ 
into the problem $[x_{1}:T_{1}] ... [x_{n}:T_{n}](F~...~x~...) = t$ 
and $x = a$ because the variables $x_{1}, ..., x_{n}$ may occur into 
the term $a$. One way to avoid this problem to use mixed prefixes
\cite{Miller}. In order that flexible-flexible equation have a 
solution we consider only unification problems such that the prefix
contains a constant in each first order type in the left of all the 
declaration of existential variables.
 
So we use a variant of the unification algorithm of \cite{Miller} such
that when one considers an equation $x = b$ of a unification problem in
a prefix $Q$, if $b$ is a term in which all the variables are first order
then if $b$ contains a constant that is declared on the right to $x$
or $b$ contains the variable $x$
one fails otherwise one substitutes to $x$ the term $b'$ obtained by
replacing in $b$ all the variables by fresh variables. Then the
problems simplifies in a sequence of equations $x = y$ where $x$ and
$y$ are first order and one substitutes the rightmost for the
leftmost. 

In the formalism of mixed prefixes, in the algorithm that transforms a
problem into a superficial problem one replaces an equation
$(F~...~a~...) = t$ into the equations $(F~...~x~...) = t$ and
$x = a$ and the prefix $Q$ into the prefix $Q[\exists x:T]$. 

\section{An Algorithm that Generalizes First Order Unification
and Second Order Matching}

{\sbd Remark:}
The flexible-flexible equations that remain in success node have
either the form 
$x = t$ (if one of the head variables is first order) or
$(f~a_{1}~...~a_{n}) = (g~b_{1}~...~b_{p})$
(if both head variables are second order). 
For equations of the first type, one can substitute the
term $t$ for $x$. If after this substitutions all the success nodes
are empty then one gets a finite complete set of unifiers.

{\sbd Remark:}
The algorithm above can be used to construct a unification algorithm
for second order terms that either gives a (possibly empty) complete
set of unifiers or fails to give any information. 

Take a second order unification problem, replace the $n$
occurrences of each second order variable by new variables $f_{1},
..., f_{n}$, apply the algorithm above. If 
\begin{itemize}
\item each success nodes of the unification tree is empty, 
\item for each success node and each second order variable $f$ of the
linear problem, $f$ is either substituted by a closed term or left
unbound
\end{itemize}
then select the success nodes such that all the closed terms substituted to the
variables $f_{1}, ..., f_{n}$ taking place of a variable $f$ of the
initial problem are equal, else fail to give any information. 

{\sbd Remark:} The algorithm above generalizes both first order
unification \cite{Robinson} and second order matching \cite{Huet76}
\cite{HueLan} as for problems in these classes it does not fail and gives a
complete set of unifiers.

\end{document}